\documentclass[twocolumn,showpacs]{revtex4}
\usepackage[dvips]{graphicx}
\usepackage{dcolumn}
\usepackage{amsmath}
\makeatletter
\def\btt#1{\texttt{\@backslashchar#1}}%
\DeclareRobustCommand\bblash{\btt{\@backslashchar}}%
\makeatother

\newcommand{\bra}{\left\langle}
\newcommand{\ket}{\right\rangle}

\begin{document}
\title{Spatiotemporal Behavior of Void Collapse in Shocked Solids}
\author{Takahiro Hatano}
\affiliation{
Center for Promotion of Computational Science and Engineering, 
Japan Atomic Energy Research Institute, Ibaraki 319-1195, Japan}
\date{\today}

\begin{abstract}
Molecular dynamics simulations on a three dimensional defective 
Lennard-Jones solid containing a void are performed 
in order to investigate detailed properties of hot spot generation.
In addition to the temperature, I monitor the number of energetically colliding particles 
per unit volume which characterizes the intensity of shock-enhanced chemistry.
The quantity is found to saturate for nanoscale voids and 
to be maximized after voids have completely collapsed.
It makes an apparent comparison to the temperature which requires much larger void 
for the enhancement and becomes maximum during the early stage of the collapse.
It is also found that the average velocity and the temperature of ejected molecules 
inside a cubic void are enhanced during the collapse because of the focusing 
of momentum and energy towards the center line of a void.
\end{abstract}

\maketitle

In liquid or solid explosives, it is known that heterogeneities 
such as bubbles or voids decrease a threshold shock strength 
above which a detonation takes place. 
When a shock passes through such defects, 
they form locally excited regions which are reffered to as ``hot spots'' \cite{dear}.
Since chemical reactions which turn shocks to detonations are believed to 
initiate inside such hot spots, the mechanism for the generation of hot spots 
is important for the understanding of detonations \cite{cheret}.
Since shock waves and void collapse are extremely fast microscopic phenomena, 
molecular dynamics simulation may be a powerful tool 
for the study of detailed properties of hot spot initiation.
As well as the studies which treat perfect crystals 
using empirical reactive potentials \cite{brenner}, 
there are a few works on the interaction between shocks and defects 
which can yield hot spots \cite{maffre,phillips,phillips2,mintmire}.
For example, Phillips speculated that chemistry would occur 
in a void below a detonation threshold \cite{phillips2}.
However, it is yet to be understood what is essential to the initiation of hot spots.
Holian et al. recently studied a two dimensional Lennard-Jones solid 
with a planar gap which might be regarded as a gap of infinite transverse width 
\cite{holian}.
They insisted that the essential ingredients to hot spot generation are:
(i) Enough shock strength for the ejection (spallation) of solid molecules into a void.
(ii) Compression of ejected particles by the far side of a void.
They estimated the overshooting temperature via a straightforward model 
constructed on the basis of the above conjectures.
However, since their estimation is not quantatively satisfactory 
(about $3/2$ of those obtained by the simulations), 
the validity of their proposal is not apparent.
Moreover, since void collapse is so fast that particles ejected into a void 
may be far from local equilibrium, it is no longer obvious that 
the ``temperature'' could be a fair index for chemistry.
We have to monitor other quantities to represent the degree of enhanced chemistry 
in this situation.

In this Letter, I focus on a dynamical aspect of hot spot generation 
to understand what is essential to shock-induced chemistry inside a void.
Especially, various spatial profiles of the ejected particles are monitored 
in order to see how the void collapse results in 
energetic intermolecular collisions which are responsible for chemistry.
As for the model, I adopt a three dimensional fcc crystal 
whose intermolecular potential is the Lennard-Jones one; 
\begin{equation}
U(r)=4\epsilon\left[
\left(\frac{\sigma}{r}\right)^{12}-\left(\frac{\sigma}{r}\right)^6
\right],
\end{equation}
whose cut-off length is set to be $4.0\sigma$.
Throughout this Letter, $\sigma$ and $\epsilon$ are set to unity (the LJ unit).
Although one might wonder about the use of nonreactive potential 
to this kind of issue, this may not be a serious limitation for the understanding 
of dynamical aspect of shock-induced chemistry; 
i.e. how the coherent (shock) energy relaxes into the thermal energy 
through the void collapse, which is the main aim of this study.
The setup of the system is as follows.
The axes of $x$, $y$, and $z$ are taken to coincide with 
$\bra 100\ket$, $\bra 010\ket$, and $\bra 001\ket$ directions, respectively.
Although there are various methods to realize shock waves in molecular dynamics simulations 
\cite{holian2}, the most popular method is adopted here that a moving piston 
of infinite mass hits the still target. 
The velocity of the piston is denoted by $u_p$.
Throughout this study, $u_p$ is fixed to be $3$ which is strong enough to cause spallation.
Shock propagates along the $\bra 001\ket$ orientation; i.e. $z$ axis.
I do not argue orientational dependence which is important to other shock-induced phenomena, 
such as plastic deformation \cite{germann} or phase transitions \cite{zhakhovskii}.
The whole system consists of $200\times 200\times 200$ unit cells.
As usual, periodic boundary conditions are applied to $x$ and $y$ directions.
A void consists of blank unit cells, the shape of which is cubic or rectangular-solid.
Each face is parallel to $\{100\}$, $\{010\}$, and $\{001\}$, respectively.
Its typical size is ranged from $5$ to $40$ unit cells for each direction.
I fix the initial temperature to be $0.1$ and the corresponding density is about $1.053$.

\begin{figure}
\includegraphics[scale=.8]{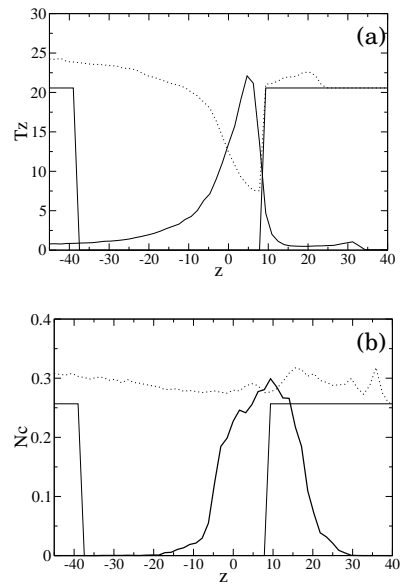}
\caption{Spatial profiles of (a) $T_z$ and (b) $N_c$ represented by thick lines 
together with the density profiles (dotted lines) which are rescaled suitably. 
(Multiplied by (a) $20$ and (b) $0.25$.)
Each graph is a snapshot at the moment of the maximization of each quantity: 
(a) $t=10.0$ for $T_z$ and (b) $t=11.0$ for $N_c$.
The initial density profiles are also imposed (thin lines).}
\label{profiles}
\end{figure}

Before proceeding to the results, I make a remark on the two kinds of averaging methods 
adopted in this study.
One is the average regarding the particles whose initial distance from 
the upstream wall of the void is less than $5$ (about 3 unit cells). 
Note that all of these particles are eventually ejected into the void 
after the passage of shock front.
I have crosschecked the simulations with different distances from $2$ to $15$, 
and the qualitative property obtained below is not influenced by a specific choice of distance.
By this method, we can trace the most energetic part of the ejected particles.
Most of the results except for Fig. \ref{profiles} are based on this averaging method.
The other method is to take spatial profiles inside a void.
This method enables us to see where and when the hot spot generation takes place.

\begin{figure}
\includegraphics[scale=.55]{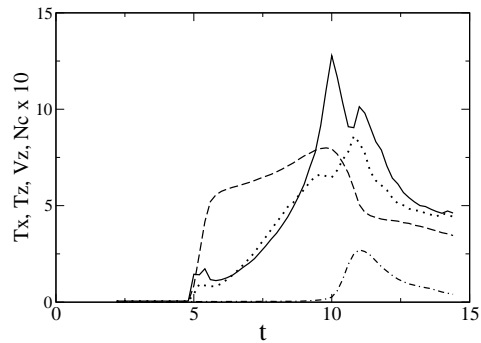}
\caption{Temporal behaviors of the longitudinal (solid line) 
and the transverse (dotted line) temperatures, the average $z$ velocity (dashed line). 
The number of energetically colliding particles which is introduced afterwards 
is also shown by the dot-dashed line (scaled by 10).
The sample molecules are those initially locate near the upstream wall of the void. 
(See text.) The void initially consists of $30\times30\times30$ blank unit cells.}
\label{Tv}
\end{figure}

Now let us turn to the results of my simulations.
First, time evolutions of the temperature during void collapse are monitored. 
However, note that the system we are considering is nonequilibrium, 
where the definitions of thermodynamic quantities including the temperature 
are precarious \cite{jou}.
Here I adopt the variance of $z$ velocity as the ``temperature'', 
which is denoted by $T_z$.
For comparison, I also monitor the variance of $x$ velocity which is denoted by $T_x$.
Note that the variances can be different values for different directions; 
e.g. $T_x$ is smaller than $T_z$ in this situation.
Temporal behaviors of $T_z$, $T_x$, and the average $z$ velocity 
denoted by $v_z$ are shown in Fig.\ref{Tv}. 
Spallation occurs at $t=5$ where $v_z$ suddenly increases 
up to about $2u_p =6$ (the spall velocity).
After the spallation, all of the quantities increase gradually until 
the ejected molecules collide with the far side of the void (approximately $t=10$). 
It should be stressed that $v_z$ increases up to about $2.6u_p$, 
which means that there is some kind of momentum transfer into the ejecta gas.
(We see this acceleration by observing the velocity distribution functions 
in the next paragraph.)
Also it should be remarked that the maxima of $T_z$ and $v_z$ 
are realized simulataneously, which is just before the collision of ejecta gas 
and the downstream wall of the void. 
It is not only the compression by the void wall that is responsible for heating of the ejecta gas 
but also some kind of focusing of momentum and energy.
However, recall that the above results only concern with the sample molecules 
which constitute the proceeding part of ejecta gas.
As for the whole gas, observation of the spatial distributions 
inside the void is needed, which we will see later in detail.

\begin{figure}
\includegraphics[scale=.32]{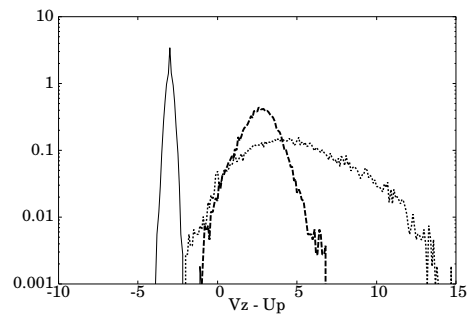}
\caption{Velocity distributions of $v_z$.
Time ordering is: (i) solid lines ($t=0$; before the spallation), 
(ii) dashed lines ($t=6.0$; just after the spallation), 
and (iii) dotted lines ($t=9.5$; just before the collision with the far side of the void).
Note that $v_z$ is shifted by $-u_p$.
All of the parameters including the sample molecules are the same as in Fig. \ref{Tv}.}
\label{vdf}
\end{figure}

\begin{figure}
\includegraphics[scale=.85]{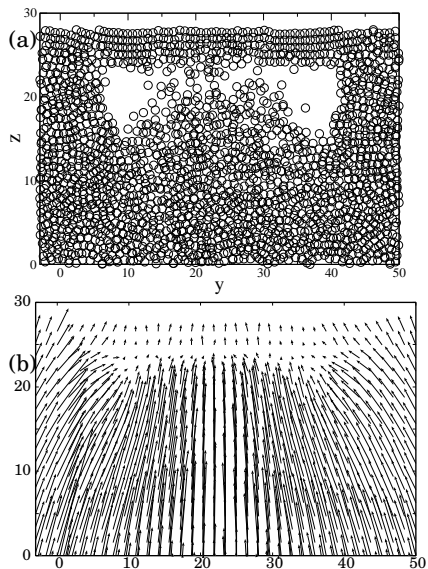}
\caption{(a) A slice-cut view (parallel to $\{100\}$ plane) of the collapsing void and 
(b) its velocity vector field. ($t=9.8$.) A shock propagates upwards.}
\label{slice}
\end{figure}

To see the above circumstances regarding the acceleration of $v_z$ in detail, 
velocity distribution function is shown in Fig. \ref{vdf}.
Strong displacement from the Maxwellian are seen after the passage of the shock front.
Note that a distribution function of $v_x$ is also observed to show 
an exponential distribution just after the spallation, 
which then eventually relaxes to the Maxwellian.
On the other hand, we can see that the distribution function of $v_z$ has 
a widely stretched tail for the positive direction, 
which develops until the collision with the far side of the void.
Note that the tail by far exceeds the spall velocity.
It is obviously responsible for the increase of $v_z$, 
since the tail develops only towards the positive direction.
The developing tail means that there is a momentum transfer into the ejecta gas, 
which may be attributed to the focusing.
As is visualized in Fig. \ref{slice}, since the shock pressure also pushes 
the side wall of the void to form a parabolic surface, 
the momenta of the collapsing wall can focus into the center line of a void, 
which eventually causes jetting and accelerates $v_z$.
Note that the acceleration during the flight does not happen in a rectangular-solid void 
with sufficiently large longitudinal length where the parabolic surface is not formed.
It is obvious that larger $v_z$ yields more energetic collision, 
which eventually leads to higher overshooting temperature inside a hot spot.
Phillips et al. \cite{phillips} also observed deviations from equilibrium distributions 
in a collapsing void using a two dimensional LJ crystal.
However, their observation focused only on a relaxation to equilibrium, 
whereas the transition from equilibrium to nonequilibrium is traced here.
Also, the positive tail which prevails spall velocity are not seen there, 
which is the original contribution of this work.
%

Apart from the temperature and the spall velocity, 
we then turn to the enhancement of chemical reactions 
which is our original concern regarding hot spots.
Since ejected particles are strongly nonequilibrium as we have seen above, 
it is no longer expected that the ``temperature'' $T_z$ can be 
a fair index of chemistry.
Although chemistry involves many intramolecular degrees of freedom, 
strong intermolecular collision is essential to the excitation of those degrees of freedom.
Therefore it is quite natural and direct to count intermolecular collisions 
for the estimation of chemistry.
In this study, I count the couple of particles whose distance is 
smaller than a certain length, $r_c$.
Then the number of the colliding particles is normalized by an initial volume of a void.
Hereafter this intensive quantity is denoted by $N_c$.
The value of $r_c$ can be various according to the molecules we are to consider; 
in this case, I choose $r_c=0.88$ which corresponds to $U(r_c)=2.5$.
Note that $U(r_c)$ gives an energy threshold for a reaction to occur.
However, since LJ potential is adopted here, we do not have the quantitative validity 
for the choice of $r_c$. 
Therefore, I crosscheck the behaviors of $N_c$ for several values of $r_c$ and 
no qualitative difference is seen regarding the results obtained in this Letter.
Typical temporal behavior of $N_c$ is shown in Fig. \ref{Tv}.
Note that $N_c$ is zero until the void collapse, which means that 
strong intermolecular collisions take place only in the void.
Also it should be remarked that the peaks of $N_c$ and of $T_z$ are not simultaneous; 
It takes a certain amount of time that $N_c$ becomes maximum after the peak of $T_z$.
Indeed, this delay means that enhanced chemistry needs higher density than overheating; 
Namely, the delay is a relaxation time for the density to increase in the collapsing void.
To confirm this idea, it is observed that the delay becomes longer for larger voids.
This situation becomes clear by monitoring spatiotemporal behavior 
of the density profile inside the void.
Fig .\ref{profiles} shows the spatial profiles of $N_c$, $T_z$, and the density, 
through which we can see in datail where and how the energetic molecular collisions 
and the overheating take place.
Note that these two snapshots are taken at the moments 
when $T_z$ and $N_c$ become maxima, respectively.
The maximum $N_c$ is realized after the whole region of the void has collapsed, 
while the preceeding maximum temperature takes place in the relatively dilute region 
on the far side of the collapsing void.
It is important to notice that $N_c$ becomes maximum after the void has totally collapsed, 
whereas $T_z$ is maximized at the beginning of collapse.
The total collapse yields high density to enhance energetic intermolecular collisions.
Therefore the time lag between the maxima of $T_z$ and $N_c$ 
are the time needed for the total collapse of the void.

The discussions so far finish spatiotemporal behaviors of the ejecta gas 
and enhanced chemistry.
As the final part of my results, I show how the size and the shape of voids affect chemistry.
Although voids in real materials can take various forms, 
it may be decomposed into two essential ingredients: 
the longitudinal length and the normal cross section.
In order to specify which component is important to the initiation of chemical reactions, 
I check rectangular-solid voids changing the transverse and the longitudinal lengths independently.
First I show the longitudinal length dependence with the transverse length 
(and hence the cross section) fixed.
We can see that $N_c$ decreases for larger longitudinal length, whereas $T_z$ increases.
As for $T_z$, this result is consistent with \cite{holian} 
where a considerable longitudinal length was needed for the enhancement of temperature.
However, the decrease of $N_c$ again shows that the temperature cannot be 
a reliable index for enhanced chemistry in this nonequilibrium situation.
\begin{figure}
\includegraphics[scale=.85]{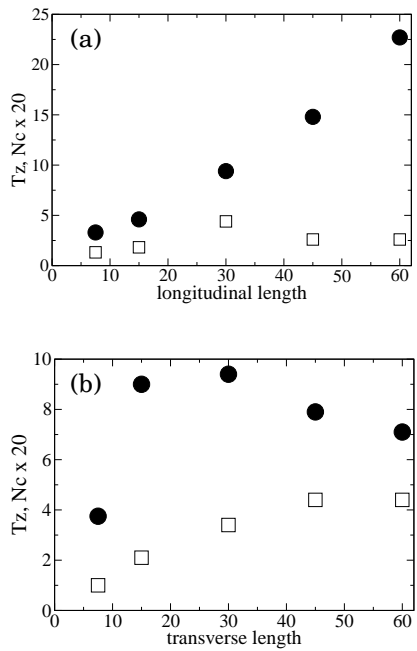}
\caption{(a) The longitudinal and (b) the transverse lengths dependences 
of $T_z$ (filled circles) and $N_c$ (blank squares, multiplied by $20$).
The length in the fixed direction of the void is $30$ in each case.}
\label{dependence}
\end{figure}
On the other hand, the cross section dependence of $T_z$ and $N_c$ seems 
quite different from the longitudinal length dependence.
The temperature $T_z$ is maximized when the void is cubic, 
whereas $N_c$ increases to saturate for the larger cross section.
It should be stressed that the cross section of $45\times 45$ 
(approximately $30\times 30$ unit cells) is enough for the saturation.
The maximization of $T_z$ with a cubic void may be a result of focusing 
discussed in the context of the acceleration of $v_z$.
Also note that a larger cubic void can yield higher temperature and $v_z$.
The saturation of $N_c$ for larger cross section means that 
the focusing in a cubic void does not seriously influence 
the number of energetic intermolecular collisions.
Recalling that higher density (the total collapse) is also required for 
the increase of $N_c$, it is plausible that the intensity of void collapse 
is not affected by the focusing, whereas the foregoing part of ejecta gas 
is accelerated to enhance the temperature.
Hence it is concluded that $N_c$ is rather insensitive to the shapes of voids, 
whereas $T_z$ has the apparent tendencies.
In any case, nanoscale voids (approximately $30$ unit cells in each dimension) 
is enough for the enhancement of energetic intermolecular collisions.

To summarize, through the observations of spatiotemporal evolutions of 
temperature and density inside a void, we have seen that energetic intermolecular collisions 
are most likely to occur when a void is totally collapsed.
This is not simultaneous with the maximization of temperature 
which is realized at the beginning of the collapse.
This time lag is due to the difference between the optimum densities 
where the each quantity is maximized.
The shape of void affects the overshooting temperature, but cannot seriously affect 
the number of energetically colliding particles per volume.
Especially, the intensity of enhanced chemistry is found to saturate for nanoscale void 
whereas the enhancement of temperature needs much larger void. 
Since chemical reactions originate from intermolecular collisions, 
monitoring only the temperature is not enough to explore 
the various properties of hot spot generation, 
such as the void-shape dependence on the enhanced chemistry.
Also, velocity distribution functions are found to deviate from the Maxwellian.
These strong nonequilibrium effects must be taken into account for 
the analysis of spallation phenomena.

The author is grateful to H. Kaburaki and F. Shimizu 
for helpful discussions and encouragements.

\end{document}